\documentclass{iaus}
\usepackage{graphicx}
\usepackage{natbib}

\title[Stellar mass ejections] 
{Stellar mass ejections}

\author[M. Jardine, J.-F. Donati, S.G. Gregory]   
{Moira Jardine$^1$ and Jean-Francois Donati$^2$ and Scott G. Gregory$^3$}

\affiliation{$^1$SUPA, School of Physics and Astronomy, University of St Andrews, North Haugh, St Andrews, KY16 9SS, UK \\ email: {\tt mmj@st-andrews.ac.uk} \\[\affilskip]
$^2$LATT, CNRS--UMR 5572, Obs.\ Midi-Pyr\'en\'ees, 14 Av.\ E.~Belin, F--31400 Toulouse, France \\email: {\tt donati@ast.obs-mip.fr} \\[\affilskip]
$^3$SUPA, School of Physics and Astronomy, University of St Andrews, North Haugh, St Andrews, KY16 9SS, UK \\ email: {\tt sg64@st-andrews.ac.uk}} 

\pubyear{2008}
\volume{257}  
\pagerange{119--126}
\jname{Universal Heliophysical Processes}
\editors{A.C. Editor, B.D. Editor \& C.E. Editor, eds.}
\begin{document}

\maketitle

\begin{abstract}
It has been known for some time now that rapidly-rotating solar-like
stars possess the stellar equivalent of solar prominences. These
may be three orders of magnitude more massive than their solar
counterparts, and their ejection from the star may form a significant
contribution to the loss of angular momentum and mass in the
stellar wind. In addition, their number and distribution provide 
valuable clues as to the structure of the stellar corona and hence
to the nature of magnetic activity in other stars.

Until recently, these "slingshot prominences" had only been
observed in mature stars, but their recent detection in an
extremely young star suggests that they may be more widespread
than previously thought. In this review I will summarise our
current understanding of these stellar prominences,  their ejection 
from their stars and their role in elucidating the (sometimes
very non-solar) behaviour of stellar magnetic fields.
\end{abstract}

\firstsection 
\section{Introduction}

During this Symposium we have learned a great deal about the Sun and its influence on its environment, but in this talk I want to begin by addressing the question ``How typical is the Sun as a star?'', Stars on the main sequence (i.e. those stars that have settled into the longest phase of their lives, when they are burning Hydrogen in their cores) can have very different interior structures depending on their mass, yet magnetic activity is almost ubiquitous among them. High mass stars have a convectively stable (or radiative) outer envelope, so how do they generate their magnetic fields? They can do this in their convective cores, although this raises the question of how to transport the flux to the surface \citep{charbonneau_macgregor_dynamo_01,brun_dynamo_05}. They can also generate magnetic fields in the radiative zone, but a very non-solar dynamo process \citep{spruit_dynamo_02,tout_pringle_dynamo_95,macdonald_mullan_dynamo_04,mullan_macdonald_dynamo_05,maeder_meynet_dynamo_05}, Alternatively, the fields may be fossils, left over from the early stages of the formation of the star \citep{moss_review_01,braithwaite_spruit_fossilfields_04,braithwaite_nordlund_dynamo_06}. Very low mass stars also have an internal structure that is very different from that of the Sun in that convection may extend throughout their interiors. In the absence of a tachocline, these stars cannot support a solar-like interface dynamo, yet they, like the high mass stars, exhibit observable magnetic fields. The mechanism by which they generate these magnetic fields has received a great deal of attention recently. While a decade or so ago, it was believed that these stars could only generate small-scale magnetic fields  \citep{durney_turb_dynamo_93, cattaneo_dynamo_99}, more recent studies have suggested that large scale fields may be generated. These models differ, however, in their predictions for the form of this field and the associated latitudinal differential rotation. They predict that the fields should be either axisymmetric with pronounced differential rotation \citep{dobler_dynamos_06}, non-axisymmtric with minimal differential rotation \citep{kuker_rudiger_97,kuker_rudiger_99,chabrier_kuker_06} or, in a very recent model, axisymmetric with negligible differential rotation \citep{browning_dynamo_08}.
\begin{figure*}[t]
\begin{center}

 \includegraphics[width=3.4in]{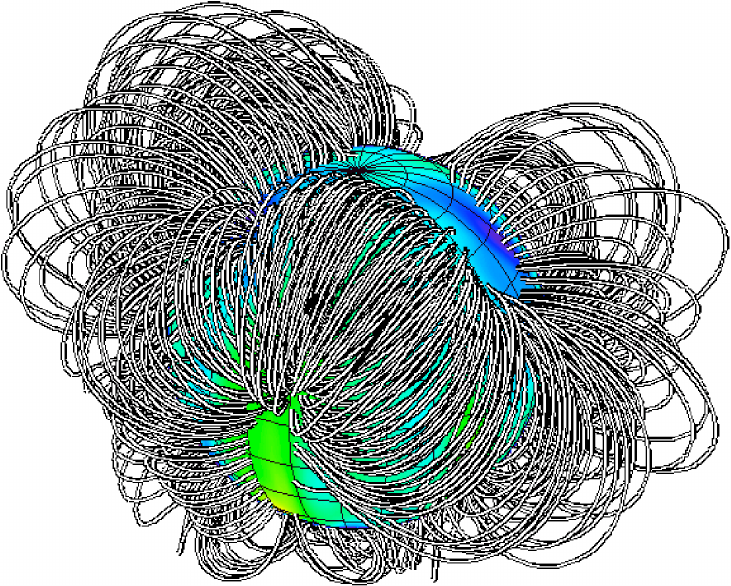} 
  \includegraphics[width=3.4in]{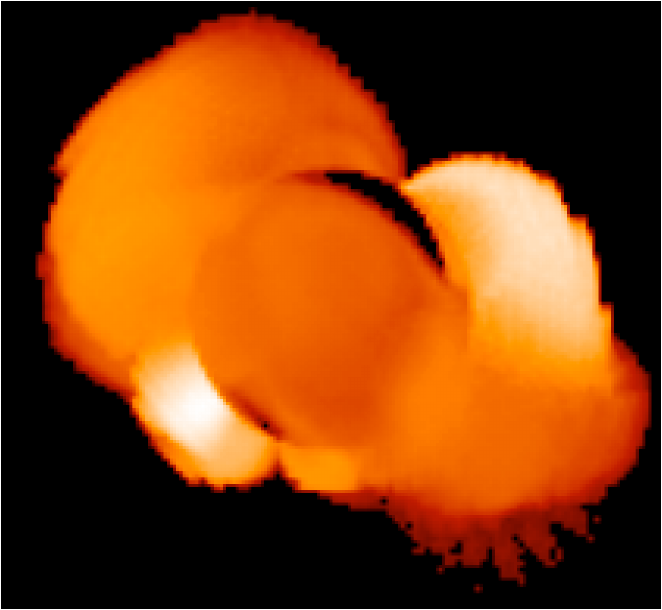} 

 \caption{Closed field lines (top) and corresponding X-ray image (bottom) for the rapidly-rotating star LQ Hya. A coronal temperature of 10$^6$K is assumed.}
   \label{fig1}
\end{center}
\end{figure*}

\section{Observing stellar prominences and magnetic fields}
With this bewildering array of magnetic field geometries, the nature of any prominences that might be confined in and ultimately ejected from these coronae becomes even more interesting, but detecting their presence is not an easy task. They can, however, be observed in rapidly-rotating stars as transient H$\alpha$ absorption features \citep{cameron89cloud,cameron89eject,cameron92alpper,jeffries93,byrne96hkaqr,eibe98re1816,barnes20PZTel,donati20RXJ}. In many instances these features re-appear on subsequent stellar rotations, often with some change in the time taken to travel through the line profile. These features are interpreted as arising from the presence of clouds of cool, dense gas co-rotating with the star and confined within its outer atmosphere. As many as six may be present in the observable hemisphere.  What is most surprising about them is their location, which is inferred from the  time taken for the absorption features to travel through the line profile. Values of several stellar radii from the stellar rotation axis are typically found, suggesting that the confinement of these clouds is enforced out to very large distances. Indeed the preferred location of these prominences appears to be at or beyond the equatorial stellar co-rotation radius, where the inward pull of gravity is exactly balanced by the outward pull of centrifugal forces. Beyond this point, the effective gravity (including the centrifugal acceleration) points outwards and the presence of a restraining force, such as the tension in a closed magnetic loop, is required to hold the prominence in place against centrifugal ejection. The presence of these prominences therefore immediately requires that the star have many closed loop systems that extend out for many stellar radii. Maps of the surface brightness distributions of these stars can be obtained by Doppler imaging, while magnetograms are now almost routinely possible with Zeeman-Doppler imaging. These maps typically show a complex distribution of surface spots that is often very different from that of the Sun, with spots and mixed polarity flux elements extending over all latitudes up to the pole \citep{donati97abdor95,donati99abdor96,strassmeier96table}. 
\begin{figure*}
\begin{center}

\includegraphics[width=2.5in]{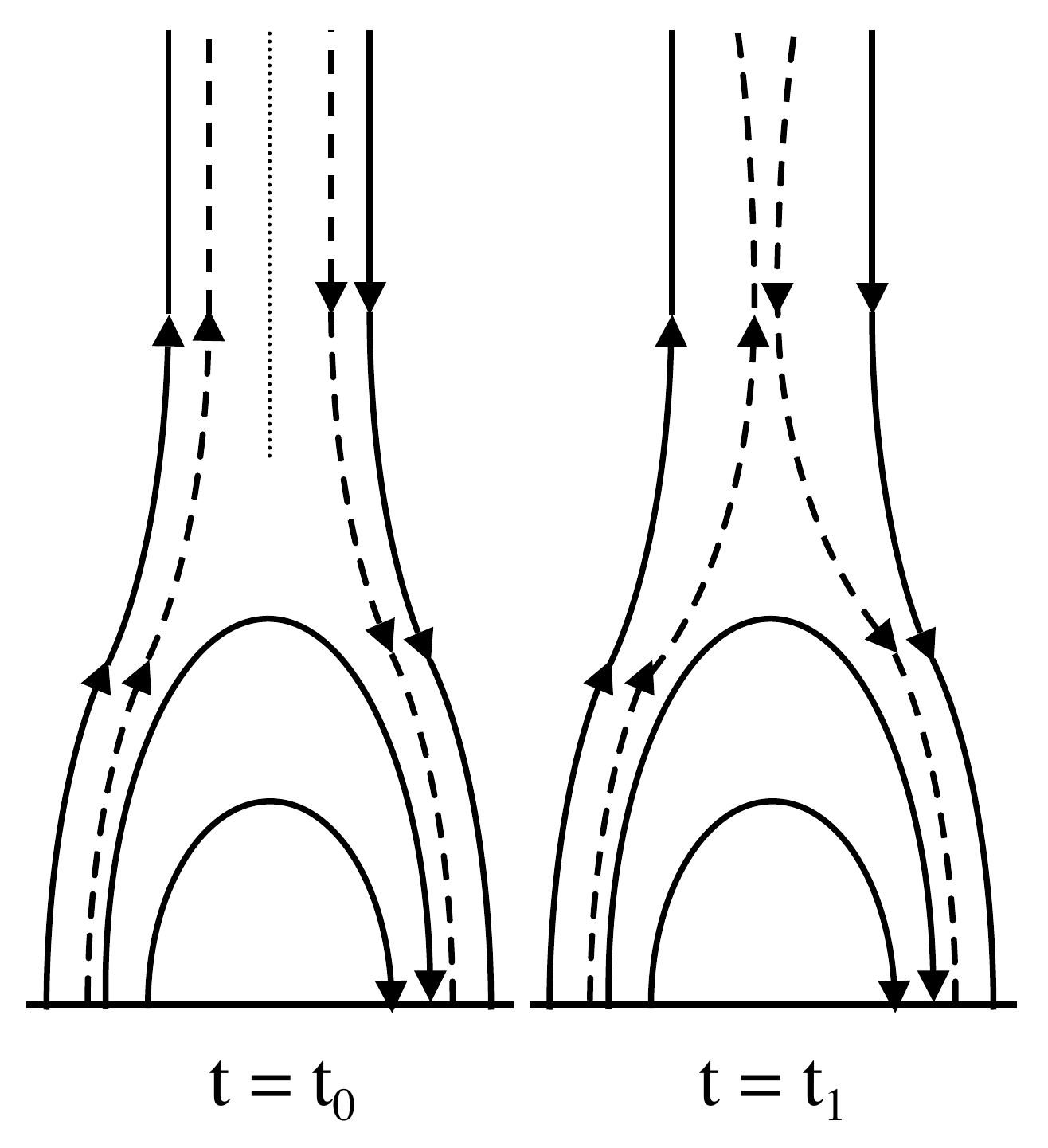}
\includegraphics[width=2.5in]{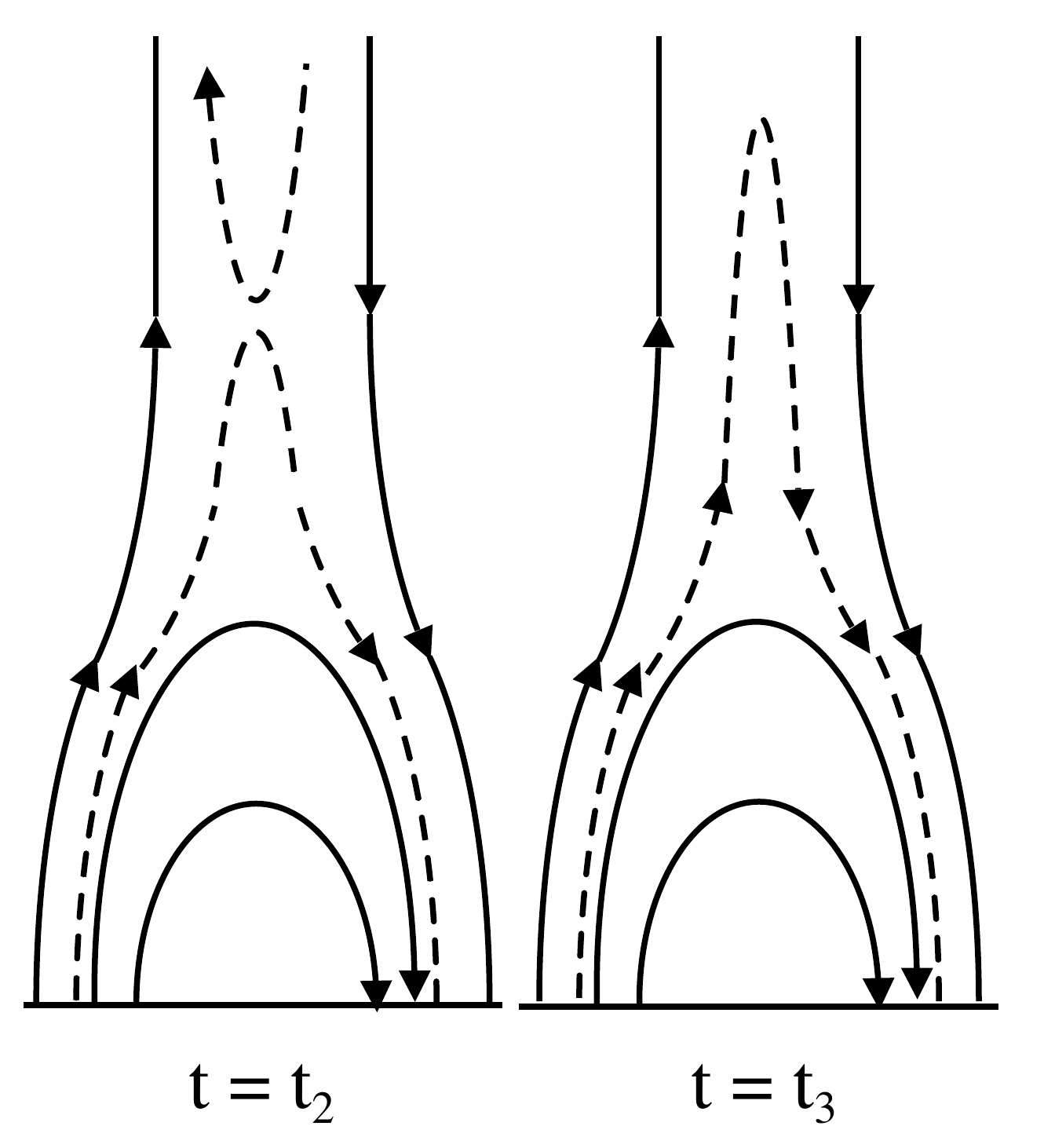}

 \caption{A schematic diagram of the formation of prominence-bearing loops. Initially, at $t=t_0$ a current sheet is present above the cusp of a helmet streamer. Reconnection at in the current sheet at $t=t_1$ produces a closed loop at $t=t_2$. The stellar wind continues to flow until pressure balance is restored, thus increasing the density in the top of this new loop. Increased radiative losses cause the loop to cool and the change in internal pressure forces it to a new equilibrium at $t=t_3$.}

  \label{fig2}
 \end{center}
\end{figure*}

From these magnetograms we can extrapolate the coronal magnetic field using a {\it Potential Field Source Surface} method \citep{altschuler69,jardine99ccf,jardine2001eqm,jardine02structure,mcivor03polar}, or using non-potential fields \citep{donati01,hussain02nonpot}. By assuming that the gas trapped on these field lines is in isothermal, hydrostatic equilibrium, we can determine the coronal gas pressure, subject to an assumption for the gas pressure at the base of the corona. We assume that it is proportional to the magnetic pressure, i.e. $p_0 \propto B_0^2$, where the constant of proportionality is determined by comparison with X-ray emission measures \citep{jardine02xray,jardine_TTS_06,gregory_rotmod_06}. For an optically thin coronal plasma, this then allows us to produce images of the X-ray emission, as shown in Fig.\ref{fig1}. This immediately highlights one of the greatest puzzles of stellar prominences:  that they are confined a such great distances - several stellar radii - that they may well be outside the extent of the closed, X-ray emitting corona. 

One way out of this problem is to confine the prominences in the wind region beyond the closed corona. \citet{jardine05proms} have produced a model for this that predicts a maximum height $y_m$ for the prominence as a function of the co-rotation radius, $y_K$ where
\begin{figure}[t]
\begin{center}

  \includegraphics[width=3.4in]{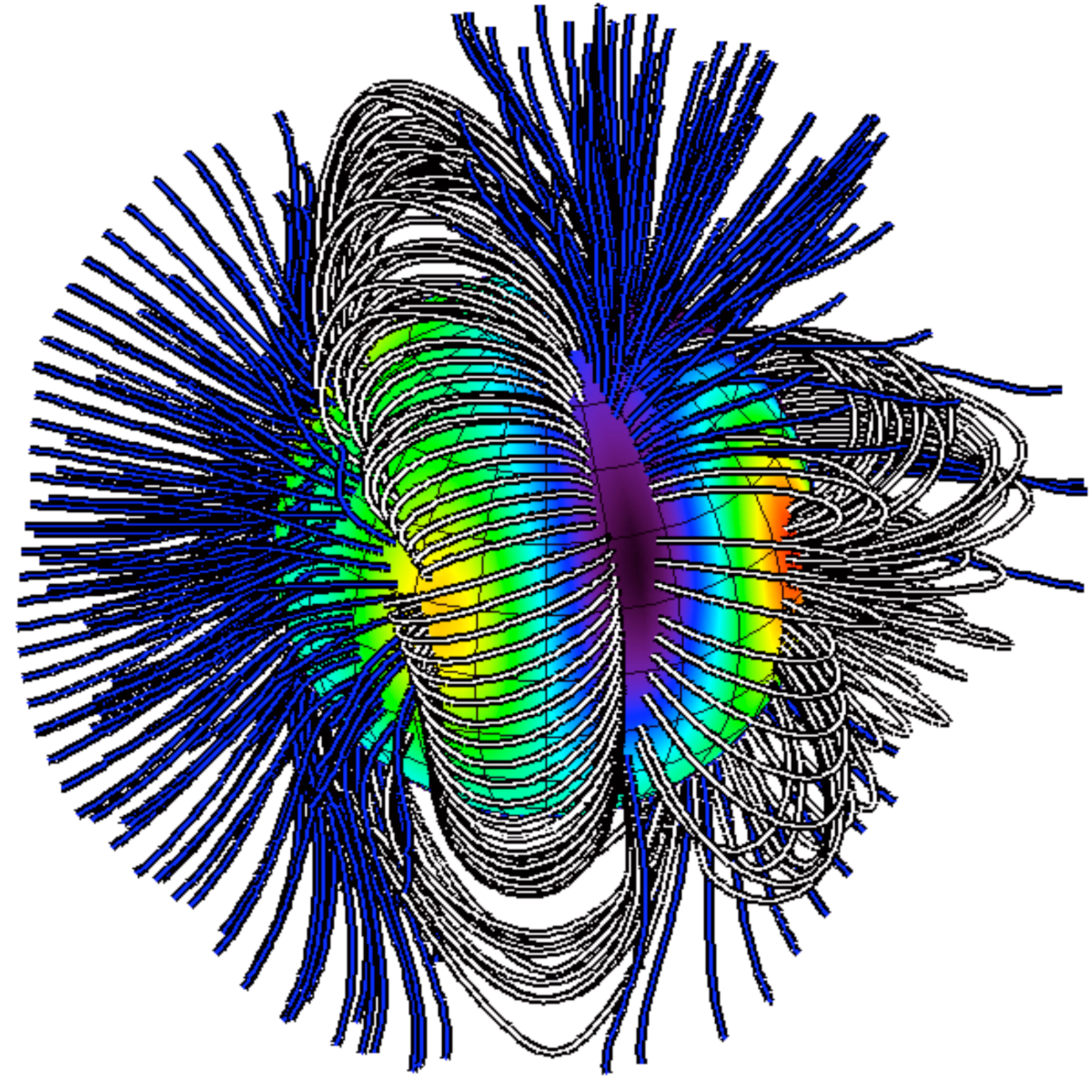} 

 \caption{Closed field lines (white) and open field lines (blue) extrapolated from a Zeeman-Doppler image of Tau Sco.}
   \label{fig3}
\end{center}
\end{figure}

\begin{eqnarray}
\frac{y_m}{R_\star} & = & \frac{1}{2}\left(
                                             -3+\sqrt{1+\frac{8GM_\star}{R_\star^3 \omega^2}}
                                                       \right) \\
                               & = & \frac{1}{2}\left(
                                             -3+\sqrt{1+8 \left[
                                                                       \frac{y_K}{R_\star} + 1
                                                                  \right]^3}
                                                       \right).
\end{eqnarray} 
Fig. \ref{fig2} shows the sequence of events that might lead to the formation of one of these ``slingshot'' prominences. The stellar wind flows along the open field lines that bound a closed field region, forming a helmet streamer. If the current sheet that forms between these oppositely-directed field lines reconnects, then a loop of magnetic field will be formed. The stellar wind will continue to flow for a short time, until pressure balance is re-established with a new field configuration. Jardine $\&$ van Ballegooijen (2006) showed that a new, cool equilibrium was possible which could reach out well beyond the co-rotation radius. The distribution of prominence heights shown in \cite{dunstone_speedymic_I_08,dunstone_speedymic_II_08} for the ultra-fast rotator Speedy Mic shows prominences forming up to (but not significantly beyond) this maximum height.
\begin{figure}[t]
\begin{center}

  \includegraphics[width=3.4in]{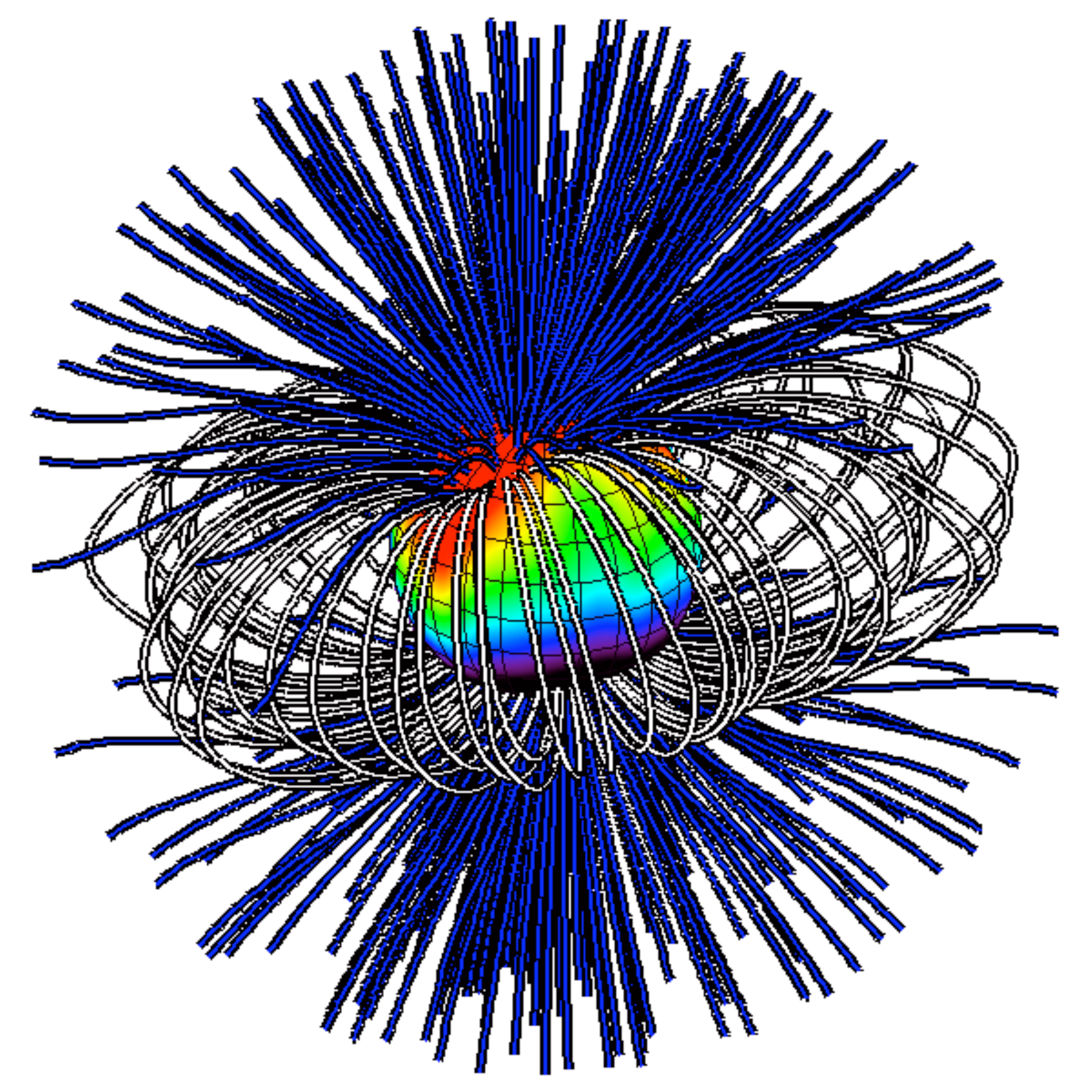} 

 \caption{Closed field lines (white) and open field lines (blue) extrapolated from a Zeeman-Doppler image of V374 Peg.}
   \label{fig4}
\end{center}
\end{figure}

\section{Stellar magnetic field variation with stellar mass and evolutionary state}
But what of the many stars whose internal structure is very different from that of the Sun? Surface magnetograms are now available for stars of a range of masses. At the high mass end, Tau Sco is a very interesting example. At 15 M$_\odot$ it has a radiative interior, and yet as shown in Fig. \ref{fig3} it displays a complex, strong field \citep{donati06tausco}. If this is a fossil field, it might be expected to be a simple dipole, but the very youth of this star, at only a million years, may be the reason why the higher-order field components have not yet decayed away. Interestingly, Tau Sco shows H$_\alpha$ absorption features that are very similar to prominence signatures in lower mass stars. In this case, however, they are attributed to a ``wind-compressed disk'' that forms when sections of the very massive wind emanating from different parts of the stellar disk collide and cool \citep{townsend_winds_05}. 
In contrast, as shown in Fig. \ref{fig4} the very low mass fully-convective star V374 Peg has a very simple, dipolar field \citep{donati06v374peg}. The highly-symmetric nature of the field and the absence of a measureable differential rotation are consistent with the recent models of Browning (2008). It is unfortunately not possible at present to detect any prominences that might be present on these very low mass stars because they stars are intrinsically too faint. Their detection would, however, be a very clear test of the magnetic structure, since in a simple dipole any prominences should, by symmetry, form in the equatorial plane.

It appears that stars with different internal structures may have dynamos that produce very different types of magnetic field. In particular, the transition from a solar-type interior to one in which the convective zone extends throughout the star appears to be associated with a decrease in field complexity. This transition happens for solar mass stars as they evolve from their very earliest stages when they are fully-convective, through the development of a radiative core as they approach the main sequence. Any associated change in the magnetic field structure is potentially very important, since the magnetic field is believed to channel the flow of material from the accretion disk that surrounds such young stars onto hotspots on the stellar surface. Significant advances have been made in the study of this {\it magnetospheric accretion} recently, with the advent of large-scale 3D MHD codes.  It appears that the structure of the magnetic field can be a crucial factor in determining the nature of the accretion \citep{rekowski_accn_04,gregory_mdot_06,long_romanova_accn_07}.
\begin{figure}[t]
\begin{center}

  \includegraphics[width=3.4in]{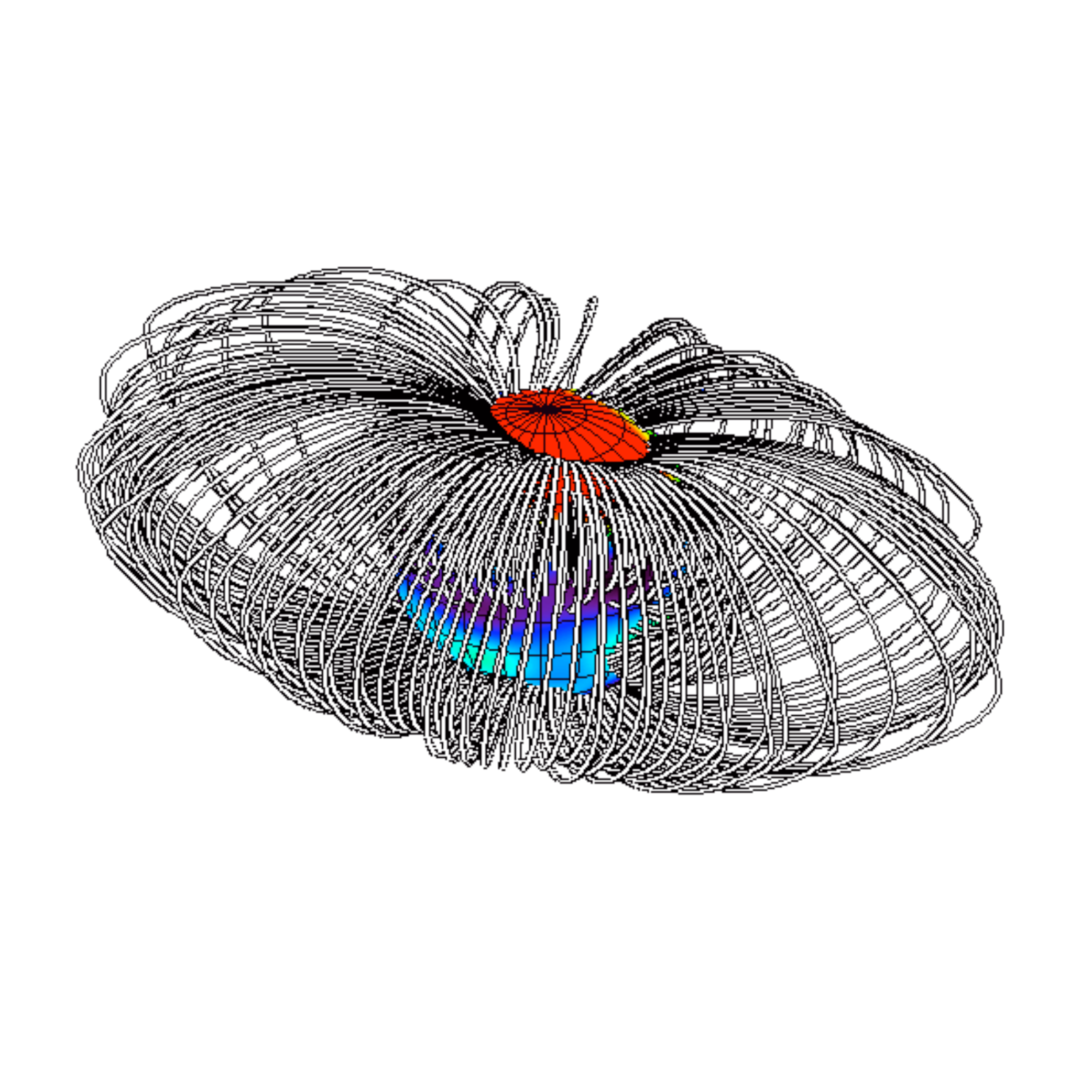} 

 \caption{Closed field lines structure extrapolated from Zeeman-Doppler image of BP Tau.}
   \label{fig5}
\end{center}
\end{figure}

It is not only the flow of material onto the star that is important, however. The loss of both mass and angular momentum in a wind is also a crucial issue since these young stars should spin up as they contract, but they are observed to have typically only moderate rotation rates. This spin-down may be achieved through the exchange of magnetic torques between the star and the disk (known as {\it disk-locking}) or through a wind \citep{konigl91,cameron_campbell_93,shu94,matt_pudritz_05}.

Determining the structure of the coronal magnetic field of these young stars is a difficult problem, however, since there are many factors than can influence it. As the stellar magnetic field drags through the disk it will be sheared and may be opened up entirely \citep{lynden_bell_boily_94}. This shearing may act to deposit energy in the corona through reconnection between the magnetic fields of the star and the disk - certainly, some of the very large flares observed in these systems may be attributed to reconnection \citep{favata_COUP_flares_05}. These processes will act in addition to the effect of the (possibly evolving) dynamo and surface flows.

\begin{figure}[t]
\begin{center}

  \includegraphics[width=3.4in]{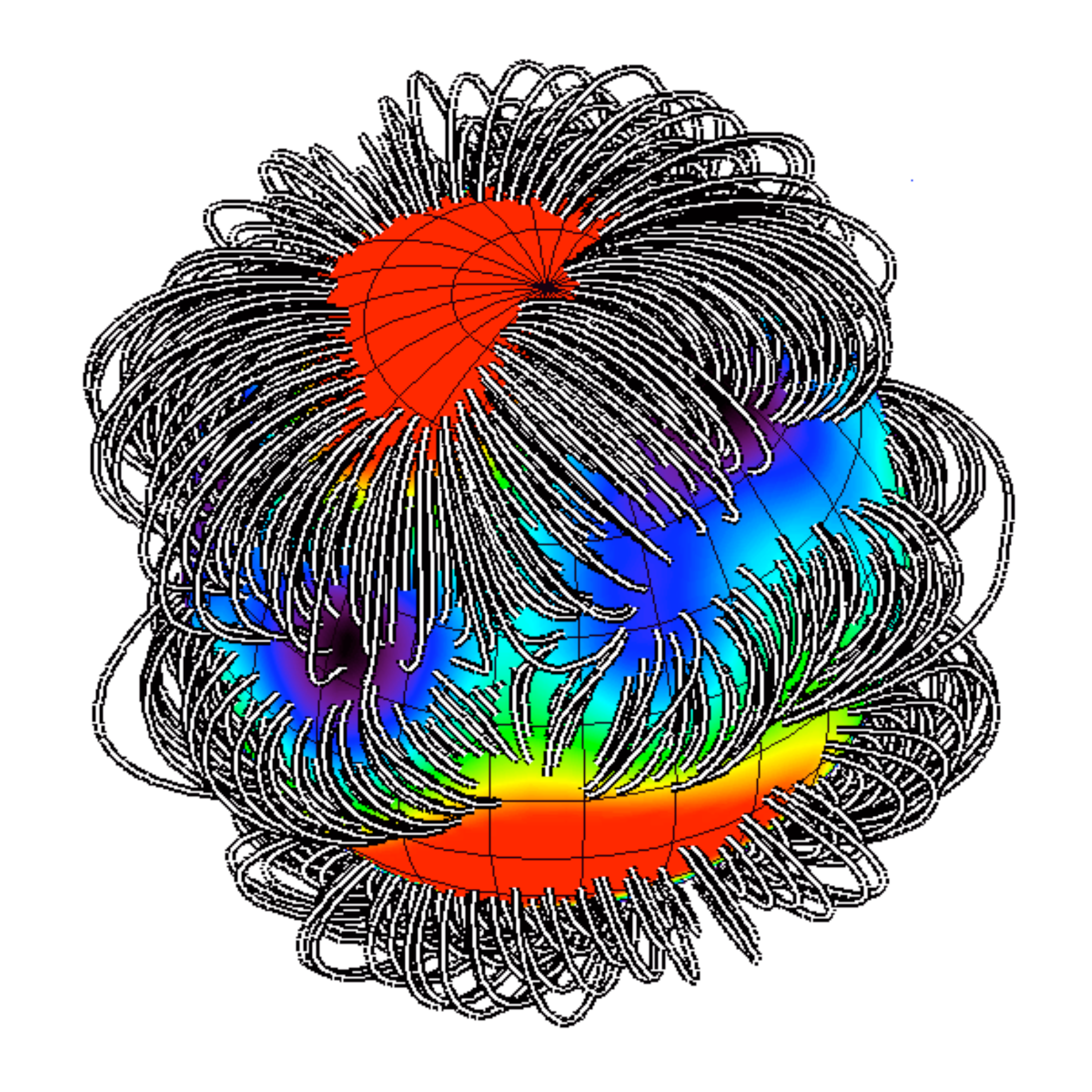} 

 \caption{Closed field lines structure extrapolated from Zeeman-Doppler image of V2129 Oph.}
   \label{fig6}
\end{center}
\end{figure}

Recently, we have successfully acquired Zeeman-Doppler images of two of these very young stars that are still accreting from their disks. One of these, BP Tau is only 0.7M$_\odot$ and is believed to be fully convective, while the other, V2129 Oph at 1.4M$_\odot$ is believed to have already developed a radiative core \citep{donati_v2129oph_07,donati_bptau_08,jardine_v2129oph_08}. As shown in Fig. \ref{fig5}, BP Tau displays a strong (1.2kG) dipolar component to its magnetic field. In contrast, as shown in Fig. \ref{fig6}, the dominant field component in V2129 Oph is the 1.2kG octupole component. The relatively stronger dipole component of BP Tau's field may allow it to carve out a larger inner hole in its disk, relative to the co-rotation radius \citep{gregory_bptau_v2129oph_08}.

\section{ Prominences in young stars}
In young stars that are still accreting it would be impossible to detect prominences, even if they were present, since the H$_\alpha$ line is so strongly affected by the accretion process that variations due to prominences could not be disentangled from those due to accretion. However, in stars that have only recently lost their disks, it is possible and indeed in one such example, TWA6, at least one prominence has been detected \citep{skelly_TWA6_08}. This is a very interesting example as the star appears to be at the boundary in its evolution between a fully convective state and the development of a radiative core. This star has a heavily-spotted surface with spots extending all the way to the rotation pole. The one prominence detected survived for at least 3 days and was situated at a radius of  4R$_\star$ - consistent with the maximum value of 4.8R$_\star$ that would be predicted by the Jardine $\&$ van Balegooijen (2006) theory.

\section{Conclusions}
It is clear from solar observations of prominences that they delineate the structure of the magnetic field, but inferring that structure from observations of prominences is not a simple (or even possible) task. This  problem is even more challenging in the case of stellar prominences where their large distance from the stellar rotation axis presents a challenge to models of their confinement by the star's magnetic field. The new Zeeman-Doppler maps of stellar surface magnetic  fields, however, show that stellar magnetic fields may be very different in stars of different mass and hence internal structure. In particular, the presence of a convectively stable core - and hence of a shear layer or tachocline separating this from the convective outer region - seems to lead to a complex, high-order field. Stars that are fully convective seem to show a much simpler structure. At present, we only have observations of prominences on mature stars with radiative cores. The low mass stars that are fully-convective are too faint to allow the detection of the transient H$_\alpha$ absorption features that are the signature of prominences. The one example we have of prominences forming in a star that is at the boundary between a fully-convective state and the formation of a radiative core is the young star TWA6 which displays a complex field. In a star with a simple dipolar field we might expect prominences to form a torus in the equatorial plane of the star - hence producing no rotational modulation. The detection of prominences in a low-mass star would be an interesting test of the field structures detected by Zeeman-Doppler methods, with potentially profound implications for dynamo theories.





\end{document}